\documentclass[pre,reprint,aps,twocolumn,superscriptaddress]{revtex4-1}

\usepackage{graphicx}
\usepackage{amssymb,latexsym,amsthm,amsmath}    
\usepackage{color}

\begin{document}

\title{Assortative mixing enhances the irreversible nature of \\ explosive synchronization in growing scale-free networks}
\author{I. Sendi\~na-Nadal}
\affiliation{Complex Systems Group, Universidad Rey Juan Carlos, 28933 M\'ostoles, Madrid, Spain}
\affiliation{Center for Biomedical Technology, Universidad
  Polit\'ecnica de Madrid, 28223 Pozuelo de Alarc\'on, Madrid, Spain}
\author{I. Leyva}
\affiliation{Complex Systems Group, Universidad Rey Juan Carlos, 28933 M\'ostoles, Madrid, Spain}
\affiliation{Center for Biomedical Technology, Universidad Polit\'ecnica de Madrid, 28223 Pozuelo de Alarc\'on, Madrid, Spain}
\author{A. Navas}
\affiliation{Center for Biomedical Technology, Universidad
  Polit\'ecnica de Madrid, 28223 Pozuelo de Alarc\'on, Madrid, Spain}
\author{J. A. Villacorta-Atienza}
\affiliation{Center for Biomedical Technology, Universidad
  Polit\'ecnica de Madrid, 28223 Pozuelo de Alarc\'on, Madrid, Spain}
\author{J.A. Almendral}
\affiliation{Complex Systems Group, Universidad Rey Juan Carlos, 28933 M\'ostoles, Madrid, Spain}
\affiliation{Center for Biomedical Technology, Universidad
  Polit\'ecnica de Madrid, 28223 Pozuelo de Alarc\'on, Madrid, Spain}
\author{Z. Wang}
\affiliation{Department of Physics, Hong Kong Baptist University,  Kowloon Tong, Hong Kong SRA, China}
\affiliation{Center for Nonlinear Studies, Beijing-Hong   Kong-Singapore Joint Center for Nonlinear and Complex Systems (Hong
  Kong) and Institute of Computational and Theoretical Studies, Hong
  Kong Baptist University, Kowloon Tong, Hong Kong}
\author{S. Boccaletti}
\affiliation{CNR- Institute of Complex Systems, Via Madonna del Piano, 10, 50019 Sesto Fiorentino, Florence, Italy}
\affiliation{Italian Embassy in Israel, 25 Hamered Street, Tel Aviv, Israel}

\begin{abstract}
We discuss the behavior of large ensembles of phase oscillators networking via scale-free topologies in the presence of a
positive correlation between the oscillators' natural frequencies and network's degrees. In particular, we show that 
the further presence of degree-degree correlation in the network structure has important
consequences on the nature of the phase transition characterizing the passage from the phase-incoherent to the
phase-coherent network's state. While high levels of positive and negative mixing consistently induce a second-order
phase transition, moderate values of assortative mixing, such as those ubiquitously characterizing social networks in the real world, 
greatly enhance the irreversible nature of explosive synchronization in growing scale-free networks. 
This latter effect corresponds to a maximization of the area and of the width of the hysteretic loop
that differentiates the forward and backward transitions to synchronization. 

\end{abstract}

\maketitle

During the last fifteen years, network theory has successfully portrayed the interaction among
the constituents of a variety of natural and man-made systems \cite{biblia,evangelio}.
It was shown that the complexity of most of the real-world networks
(RWNs) can be reproduced, in fact, by a growing process that eventually
shapes a highly heterogeneous
(scale-free, SF) topology in the graph's connectivity pattern \cite{Albert1999}.
Furthermore, such a SF degree distribution affects, on its turn, in a
non-negligible way almost all the dynamical processes taking place over RWNs \cite{biblia}.

Actually, and distinct from the degree distribution, many other
important properties account for the fine details of
the structure of any RWN, mostly due to particular forms of correlation (or mixing) among the network
vertices \cite{Newman2003}. The simplest case is degree correlation \cite{Newman2002},
in which the network constituents tend to choose their interactions according to
their respective degrees. Remarkably, non-trivial forms of degree
correlation have been (experimentally and ubiquitously) detected in
RWNs, with social networks displaying typically an assortative mixing (i.e. a situation in which
each network's unit is more likely to connect to other nodes with approximately the same degree), while
technological and biological networks exhibiting a disassortative mixing (that takes place when
connections are more frequent between vertices of fairly different degrees).
Both an assortative and a disassortative mixing are known to
considerably affect the organization of the network into collective dynamics, such as synchronization
\cite{Sorrentino2006,Chavez2006}.

Indeed, the most studied emerging collective dynamics in SF networks is certainly synchronization
\cite{biblia,ArenasPR2008}, as such a state plays a crucial role in
many relevant phenomena like, for instance, the emergence of coherent global behaviors
in both normal and abnormal brain functions~\cite{epilepsis}, the
food web dynamics in ecological systems~\cite{maccan98} or in the stable operation of electric power grids \cite{rohden2012,peng2013,Menck2014}.
In particular, it has been recently shown that the transition to the
graph's synchronous evolution may have either a reversible, or an irreversible discontinuous nature.
The former case is what traditionally investigated in coupled
oscillators, where a second-order phase transition characterizes the continuous
passage from the incoherent to the coherent state of the network
\cite{kuramoto1975,strogatz2000}.
The latter, instead, corresponds to a discontinuous transition, called explosive
synchronization (ES) \cite{Jesus2011}. Based on Kuramoto oscillators, ES has rapidly become a subject of enormous interest
\cite{Pazo2005,Jesus2011,Leyva2012,Leyva2013a,Leyva2013b,Zou2014}.
While originally it was suggested that ES was due to a positive
correlation between the natural frequencies of oscillators and the
degrees of nodes \cite{Jesus2011}, more recent studies
have proposed unifying frameworks of mean-field, where the effective
couplings are conveniently weighted \cite{Zhang2013,Leyva2013a}.
Yet, only preliminary studies exist on the effect of degree mixing on
ES \cite{Li2013,Zhu2013,Liu2013}. 

In this paper, we focus on ES of coupled phase oscillators in growing
SF networks, and show that degree mixing has important
effects on the nature of the phase transition characterizing the
passage from the phase-incoherent to the
phase-coherent network's state.
In particular, we will show that assortativity has the effect of enhancing the width and area of the hysteretic region associated to ES,
thus magnifying the irreversible nature of that transition. Actually,
our evidence is that there is an optimal level of assortativity in
growing SF networks, for which the area of hysteresis in ES is maximal.

To this purpose, let us start with considering a network of $N$ coupled phase oscillators
whose phases $\theta_i$ ($i=1,...,N$) evolve according to the Kuramoto model \cite{kuramoto1975}:

\begin{equation}
  \frac{d\theta_i}{dt}=\omega_i + \sigma\sum_{i=1}^N a_{ij} \sin(\theta_j-\theta_i),
  \label{eq:kuramoto}
\end{equation}
where $\omega_i$ is the natural frequency of the $i^{th}$ oscillator.
Oscillators interact through the sine of their phase difference, and
are coupled according to the elements of the network's adjacency matrix $a_{ij}$, being $a_{ij}=1$
if oscillators $i$ and $j$ are coupled, and $a_{ij}=0$ otherwise. The
strength of the coupling is controlled by the parameter $\sigma$, by increasing which
one eventually (i.e. above a critical value of the coupling) promotes the transition to the coherent state,
where all phases evolve in a synchronous way \cite{kuramoto1975,kuramoto1984}.

Following the changes in the level of synchronization among oscillators as the
coupling strength increases is tantamount to monitoring the classical order parameter $s(t)=
\frac{1}{N} |\sum_{j=1}^N e^{i \theta_j(t)}|$ \cite{kuramoto1975}. Indeed, the time average of
$s(t)$, $S=\langle s(t)\rangle_T$, over a large time span $T$ assumes values ranging from $S\sim 0$
(when all phases evolve independently) to
$S\sim 1$ (when oscillators are phase synchronized).

Typically, Eqs.~(\ref{eq:kuramoto}) give rise to a second-order phase
transition from $S \simeq 0$ to $S \simeq 1$ for a unimodal and even frequency distribution $g(\omega)$, with a critical coupling at
$\sigma_c=2/(\pi g(\omega=0))$ for the case of all-to-all
connected oscillators \cite{acebron2005}, and $\sigma_c'=\sigma_c \frac{\langle k\rangle}{\langle k^2\rangle}$ 
for the case of a complex network with first and second moments of the degree distribution, $\langle k\rangle$ and $\langle k^2\rangle$ respectively \cite{ArenasPR2008}. However, in the last few years it was pointed out that a different scenario (ES)
can arise, featuring an abrupt, first-order like, transition
to synchronization, and associated with the presence of an hysteretic loop
\cite{Pazo2005,Jesus2011,Leyva2012,Leyva2013a,Leyva2013b}.
In this latter case, the forward (from $S \simeq 0$ to $S \simeq 1$) and backward (from $S \simeq 1$ to $S \simeq 0$) transitions
occur in a discontinuous way and for different values of the coupling strength, this way marking
an irreversible character of the phase transition, which is of particular interest at the moment of
engineering (or controlling) magnetic-like states of synchronization \cite{Leyva2013b}.

In the following, we concentrate on ES in growing SF networks
\cite{Albert1999}, when a microscopic
relationship between the structure and the dynamical properties of the
system is imposed, and we will investigate the influence on the
nature of that transition when node degree-degree
correlations are present in the network \cite{Newman2002}.
In particular, and following the approach of Ref.~\cite{Jesus2011},
we will choose a direct proportionality between the frequency and the degree distribution ($g(\omega)=P(k)$), 
implying that each network's oscillator is assigned a natural frequency equal to its
degree, $\omega_i=k_i$, being $k_i$  the number of
neighbors of the oscillator $i$ in the network.

As for the stipulations followed in our simulations, networks  are
constructed following the procedure introduced in
Ref. \cite{GGPRE73}.  Such a technique, indeed, allows constructing graphs with the same
average connectivity $\langle k \rangle$, and grants one the option of continuously interpolating from
Erd\H{o}s-R\`enyi (ER) \cite{ER} to SF \cite{Albert1999} topologies, by tuning a single parameter $0 \leq \alpha \leq 1$.
With this method, networks are grown from an initial small clique of
size $N_0>m$, by sequentially adding nodes, up to the desired graph size
$N$. Each newly added node then establishes $m$ new links having a
probability $\alpha$ of forming them randomly with already existing vertices, and a probability
$1-\alpha$ of following a {\it preferential attachment} rule for the selection of
its connection. When the latter happens, we use a generalization of the original preferential attachment rule \cite{Albert1999} that includes an initial and constant attractiveness $A$ for each of
the network's sites, so that the attractiveness of node $i$ (the probability that such a node has to receive
a connection) is $A_i=A+k_i$ \cite{Dorogovtsev2000}.
The result of the above procedure is that the limit $\alpha=1$ induces an ER
configuration, whereas the limit $\alpha=0$ corresponds to a SF network
with degree distribution $P(k) \sim k^{-\gamma}$, with
$\gamma=2+A/m$ (when $A=m$, $\gamma=3$ and  the Barab\'asi-Albert (BA)
model is recovered).

It is well known that the BA  model does not
exhibit any form of mixing in the thermodynamic limit (${N \to \infty} $).
As for degree correlations, we quantify them using the Pearson
correlation coefficient $r$ between the degrees of all nodes at either
ends of a link, that can be calculated as in Ref. \cite{Newman2002}:

$$r=\frac{L^{-1}\sum_i j_i k_i  -[L^{-1}\sum_i\frac{1}{2}
  (j_i+k_i)]^2}{L^{-1}\sum_i \frac{1}{2}(j_i^2+k_i^2) -[L^{-1}\sum_i\frac{1}{2}
  (j_i+k_i)]^2}, $$
where $j_i$ and $k_i$ are the degrees of the nodes at the ends of
the $i$th link, with $i=1,\cdots,L$. Actually, one has that $-1\le r \le 1$,
with positive (negative) values of $r$ quantifying the level of assortative (disassortative) mixing
of the network.

In order to generate SF networks with given and tunable levels of degree mixing,
we follow the strategy described in Ref.~\cite{Newman2002}.
After the network has been grown, we choose a pair of links at
random and monitor the degrees of the four nodes at
the ends of such links. The links are then rewired in such a way that
the two largest and the two smallest degree nodes become connected
provided that none of those links already exist in the network (in
which case the rewiring step is aborted and a new pair of links is
selected). Repeating iteratively such a procedure results in progressively increasing the assortativity of the network, in that
more and more connected nodes of the network will display similar
degree. Conversely, if the rewiring is operated in a way to
determine that the largest (second
largest) and the smallest (second smallest) degree nodes are
connected, the resulting network becomes progressively dissasortative.

Therefore, we first generate a network of size $N=10^3$ with a given mean degree $\langle k\rangle=2m$
and slope $\gamma$ as described above. Then, we check whether this
network is uncorrelated, that is whether $r=0$. If not (which is always the case due to finite size effects), we perform the
link rewiring procedure until the network is neutral (i.e. with no degree
correlations). Finally, we take this network as our network of reference, and
perform the link rewiring procedure in order to produce an ensemble of networks, each one having the
 same degree distribution, but different values of the assortativity
coefficient $r$.

\begin{figure}
  \centering
  \includegraphics[width=0.35\textwidth]{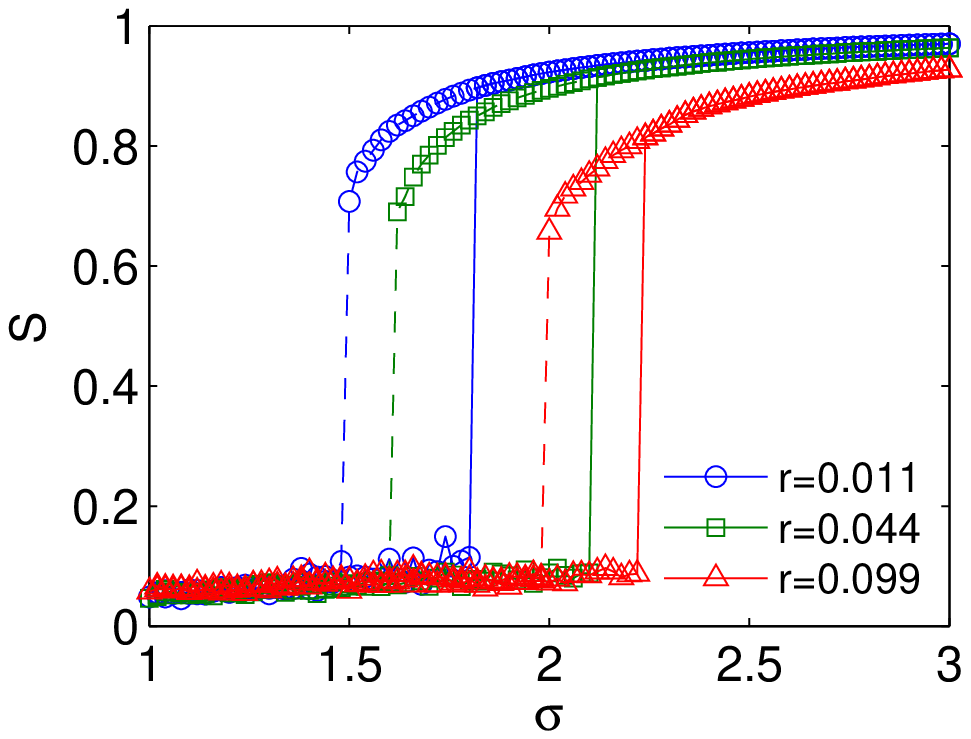}
  \includegraphics[width=0.35\textwidth]{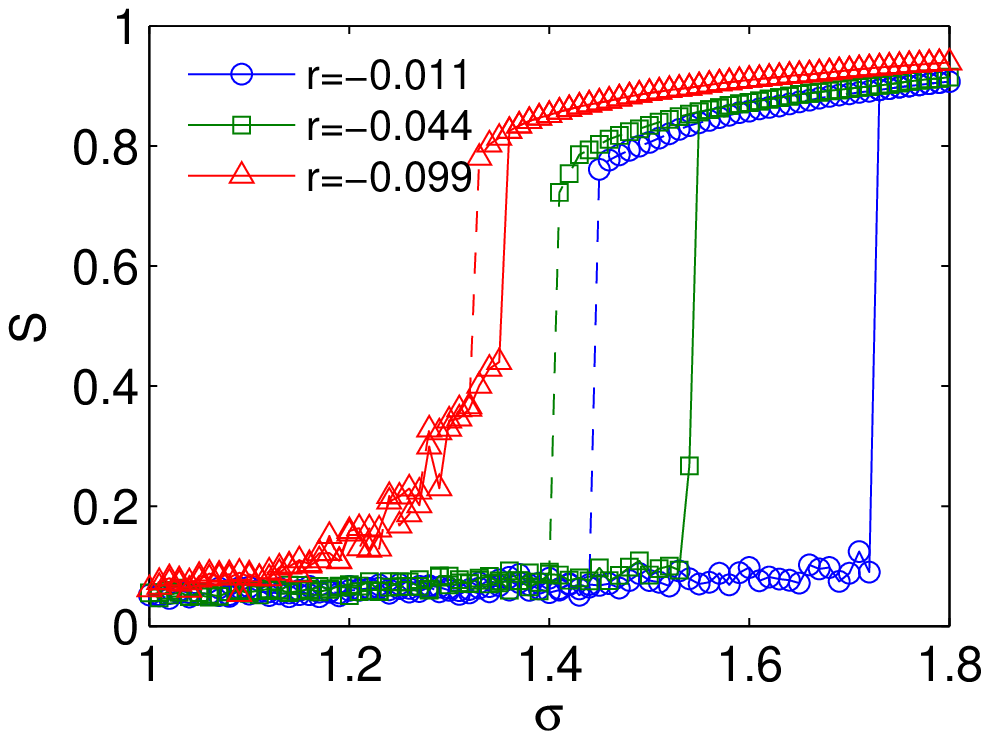}
  \caption{(Color online). Order parameter $S$ {\it vs.} $\sigma$ (see text for definition) 
  for the forward (solid curves) and backward (dashed curves) transitions. Results refer to
     SF networks displaying different levels of assortative (top panel) and
    disassortative (bottom panel) mixing. Curves are colored accordingly to the specific
    value of the parameter $r$ (reported in the legend of each panel).  In all cases, $N=10^3$,
    $\langle k \rangle =6$, $\gamma=2.4$, and natural frequencies are
    set to be $\omega_i=k_i$. \label{fig1} }
\end{figure}

\begin{figure}
  \centering
   \includegraphics[width=0.35\textwidth]{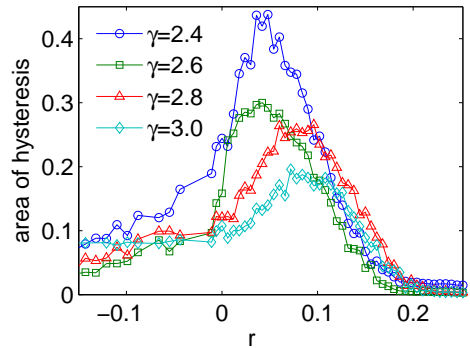}
  \caption{(Color online). Area of the hysteretic region delimited by the forward and
    backwards synchronization curves {\it vs.} the assortativity (disassortativity)
    coefficient, for different values (reported in the legend) of the exponent $\gamma$ of the
    degree distribution $P(k)\sim k^{-\gamma}$. Each point is an average over 10 different simulations, each
one corresponding to a different network realization for the given
assortative  or dissasortative  mixing level. In all cases, networks are SF
with $N=10^3$, $\langle  k\rangle =6$, and  $\omega_i=k_i$. \label{fig2}
}
\end{figure}

Figure~\ref{fig1} illustrates the effect of imposing  a degree mixing on a growing SF network.
Precisely, Eqs.(\ref{eq:kuramoto}) are simulated on top of SF networks ($N=10^3$,
    $\langle k \rangle =6$, $\gamma=2.4$) with assortative (top panel), or
disassortative (bottom panel) mixing. For each value of $r$, we monitor
the state of the network through the order parameter $S$ by gradually increasing $\sigma$ in steps $\delta
\sigma$ (forward tuning) and also in the reverse way, i.e. departing from a network
state where $S = 1$ and gradually decreasing the coupling by $\delta
\sigma$ at each step (backward tuning).
From the results shown in the bottom panel of Fig.~\ref{fig1} it is evident that an increasing level of disassortative mixing reduces
the threshold of the forward transition (which is consistent with the
general claims of Refs.~\cite{Sorrentino2006,Chavez2006} that
disassortativity favors network's synchronizability), but it
progressively reduces the hysteretic area associated with ES, up to
eventually recovering a second-order reversible transition.
At variance, and remarkably, the effects of assortativity (top panel of Fig.~\ref{fig1}) are seemingly non trivial: the threshold for the forward synchronization has an increasing trend with $r>0$, but the area of hysteresis appears to widen for intermediate values of $r$.

In order to provide a more quantitative analysis, extensive numerical simulations of Eqs.(\ref{eq:kuramoto})
were performed at various values of $r$, and for SF networks with different slopes
$\gamma$ and same mean degree $\langle k\rangle=6$. The results are summarized in Fig.~\ref{fig2}.
The most relevant
result is that the hysteresis of the phase transition is highly
enhanced (weakened) for positive (negative) values of the assortative mixing parameter, and that there
is an optimal positive value of $r$ where the irreversibility of the phase
transition is maximum. 

As the slope of the power law of the degree
distribution becomes steeper (large values of $\gamma$), the enhancement produced by a positive degree mixing gradually vanishes
and the optimum shifts to higher
values of $r$. Notice, finally, that null values of the hysteretic area indicate
that the transition has lost its irreversible character, so that degree mixing can turn
an explosive irreversible phase transition into a second order, reversible one.

\begin{figure}
  \centering
    \includegraphics[width=0.35\textwidth]{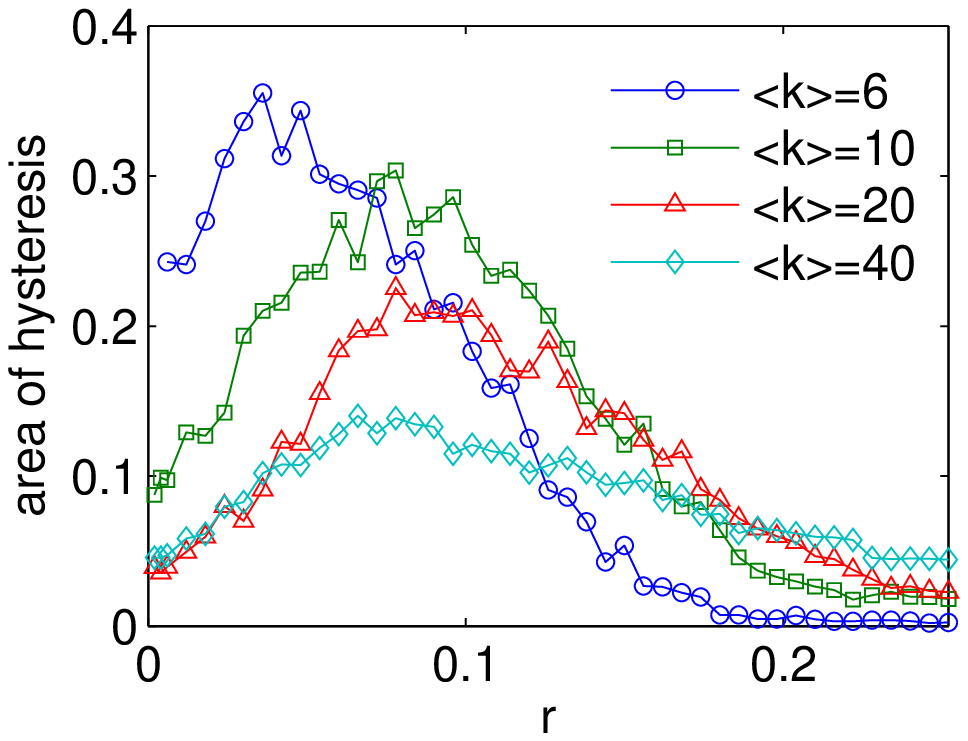}\\
    \includegraphics[width=0.35\textwidth]{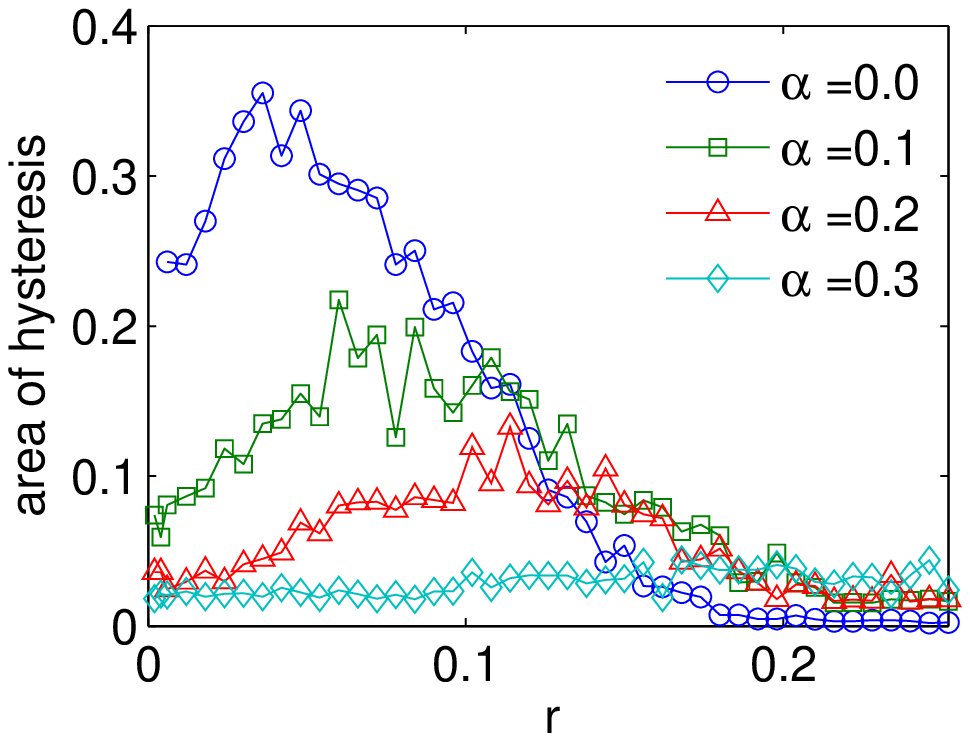}
\caption{(Color online). Area of hysteresis as a function of the mixing
  coefficient $r$. In the top panel, the different curves corresponds to different values of the mean degree $\langle k
  \rangle$ (reported in the legend), while in the bottom panel $\langle
  k\rangle=6$ and the network heterogeneity is varied by means of increasing the parameter $\alpha$ (see the legend for the color code of the
  different reported curves), from pure SF
  ($\alpha = 0$) to $\alpha=0.3$ ($\alpha=1$ corresponds to a pure ER
  network). In all cases, each point is an average of 5 simulations, $N=10^3$, $\gamma=2.4$, and $\omega_i=k_i$.
 \label{fig3} }
\end{figure}

Further information can be gathered by exploring the effect of varying the mean degree $\langle k
\rangle$, and the level of heterogeneity $\alpha$ of the network. In the top panel of Fig.~\ref{fig3} it can be
observed that already at $r=0$ (uncorrelated networks), increasing the
mean degree results in narrowing the area of hysteresis, with the consequence that the
phase transition becomes smoother and smoother, until eventually ES is lost. For generic values of $r$, as $\langle k\rangle$
increases, the curves of the area of hysteresis are attenuated and
shifted to higher values of $r$. Regarding the effect of the heterogeneity in the
network's connectivity (bottom panel of Fig.~\ref{fig3}), moving from pure SF networks ($\alpha=0$) to slightly larger
values of $\alpha$ rapidly deteriorates the enhancement of
hysteresis. However, a positive degree mixing can still turn a second
order phase transition (for $r=0$) into an abrupt and irreversible one at a value of
$r\sim 0.1$ when $\alpha=0.2$.

Finally, we draw attention to another relevant observation: our results seem to indicate
that a crucial condition to obtain a strong irreversibility
in ES is having an underlying growing process
through which the SF topology is shaped. 
To show this point, we comparatively consider ensembles of networks
displaying the very same SF distributions as those of Fig.~\ref{fig2}, but this time we
construct the SF topology by means of the so called {\it configuration
  model} (CM) \cite{CF}. Once again, we set $N=10^3$, $\langle k\rangle=6$,  and we distribute the oscillators' frequencies so as to determine a direct
correlation with the node degree ($\omega_i=k_i$).

\begin{figure}
  \centering
    \includegraphics[width=0.35\textwidth]{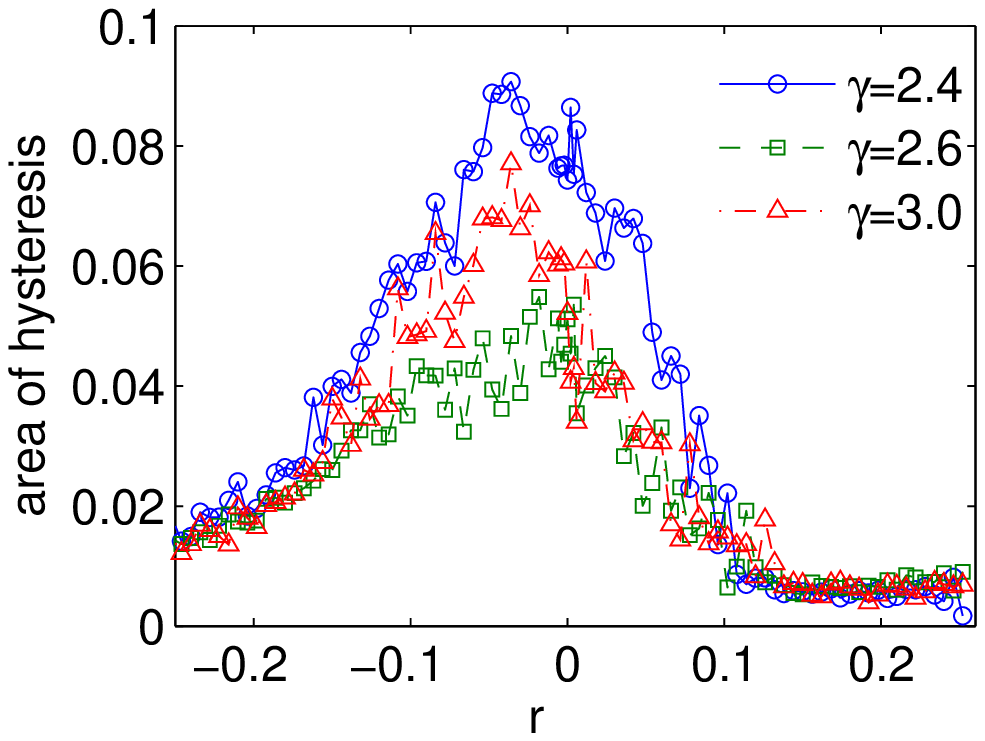}
  \includegraphics[width=0.35\textwidth]{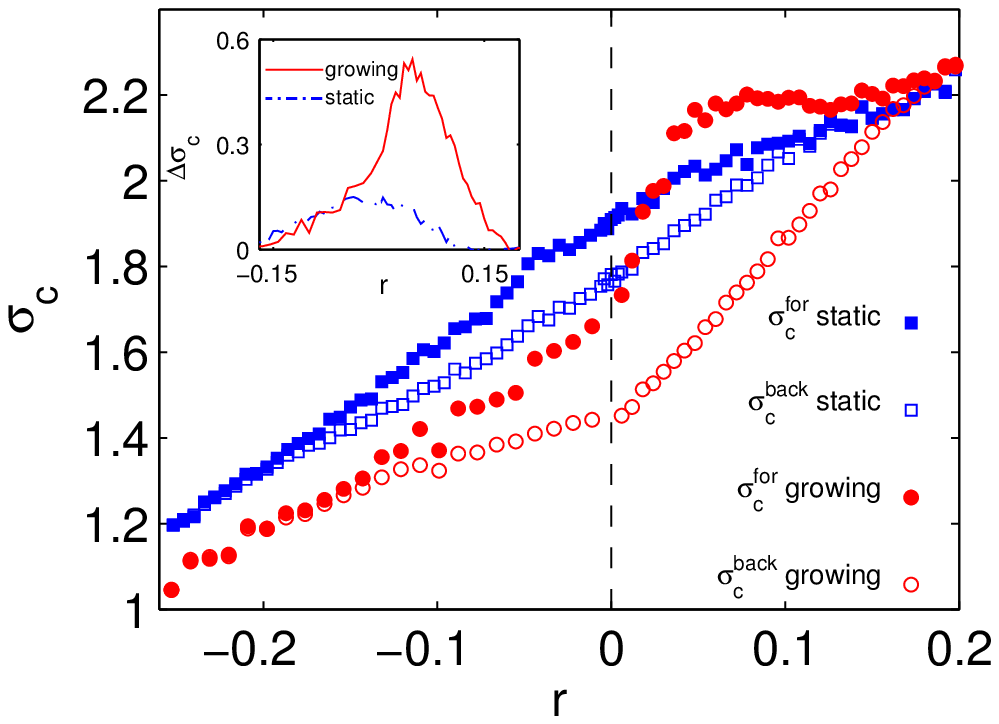}
\caption{(Color online). (Upper panel) Area of hysteresis as a
  function of the mixing  coefficient $r$ for static networks obtained
  with the configuration model and
 displaying the very same degree distributions $P(k)$ as the networks
 used in Fig.~\ref{fig2}. Each point is an ensemble average over 20 different network realizations. In all cases,
  $N=10^3$, $\langle k\rangle=6$. (Bottom panel)
Critical coupling strengths for the forward ($\sigma_{c}^{for}$, solid symbols)  and
backward ($\sigma_{c}^{back}$, hollow symbols) transition for growing (circles) and
static (squares) SF networks as a function of the degree
mixing $r$. $N=10^3$, $\langle k\rangle=6$, and $\gamma=2.4$. Vertical dashed line marks $r=0$. 
The inset of the bottom panel reports the corresponding width of the hysteresis curves, calculated as $\Delta
\sigma_c=|\sigma_{c}^{for}-\sigma_{c}^{back}|$ for growing ($\circ$) and static ($\Box$) networks.\label{fig4}
 }
\end{figure}

The results are reported in Fig.~\ref{fig4}. In the top panel, one clearly sees 
how growing SF networks present a larger hysteresis area than CM networks for any value of $r$. However, comparing the top panel of Fig.~\ref{fig4} with Fig.~\ref{fig2} allows one to realize that both networks present 
and enhancement of irreversibility in connection with an increase in the degree-degree correlation
. 
In the bottom panel of Fig.~\ref{fig4} we report the critical coupling strengths $\sigma_{c}^{for}$ and $\sigma_{c}^{back}$
characterizing the forward (solid symbols)  and
backward (hollow symbols) transitions to synchronization, respectively. 
As a function of the degree mixing, the curves obtained for growing (circles) and
static (squares) SF networks have completely different trends. For CM networks, both the forward and the backward
transitions are associated with critical coupling strengths that monotonically increase with $r$. On the contrary, for growing SF networks the curve of 
$\sigma_{c}^{for}$ displays a clear plateau for intermediate values of $r$ which occurs right in correspondence with the maximum in the area of the hysteresis observed in Fig.~\ref{fig2}. This suggests that specific topological meso-scales pronouncedly arise at those values of $r$ which influence the forward transition, having the effect of obstructing the otherwise increasing trend of $\sigma_{c}^{for}$.
The inset of the bottom panel of Fig.~\ref{fig4} reports the corresponding width of the hysteresis curves, calculated as $\Delta
\sigma_c=|\sigma_{c}^{for}-\sigma_{c}^{back}|$ for growing ($\circ$) and static  ($\Box$) networks, and underlines once again the existence of a maximum
for $\Delta \sigma_c (r)$ for growing SF networks.

\begin{figure}
  \centering
  \includegraphics[width=0.35\textwidth]{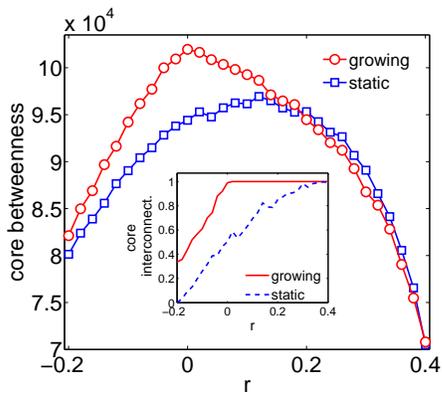}
  \caption{Betweenness centrality and interconnectivity (inset) of the core (first three higher-degree hubs) as a function of the degree mixing $r$. The interconnectivity of $n$ nodes is defined as the number of links connecting those nodes over the number of possible links $n(n-1)/2$. These characteristics have been evaluated by averaging 
for the nodes within the core over 100 realizations 
of each type of SF network, growing (red circles) and static (blue squares) with $N=10^3$, $\langle k\rangle=6$, and $\gamma=2.4$.}  \label{fig5}
\end{figure}

Figure~\ref{fig4} illustrates how the irreversibility of the ES depends on the the second order properties of the network topology, as an enhancement of the hysteresis appears to be associated with a moderate increase in the degree-degree correlation, recovering a second order transition for large values of positive and negative $r$. 
This complex behavior can be understood by examining the inner mechanism of the frequency-degree correlation. 
Explosive transitions result from a frustration in the path to synchronization \cite{navasprep}. 
In the case of ER networks, where the path to synchronization starts from multiple seeds homogeneously distributed 
in the network, this frustration can be induced by imposing a gap in the frequency differences of each pair of nodes. 
The larger the gap frequency, the higher the frustration (explosivity) of the system, which shows a positive correlation 
between the explosive character of the system and the width of the hysteresis \cite{Leyva2013b}. 
In the case of general SF networks, this path starts from the hubs, leading to the synchronization of the system 
by progressively recruiting nodes  \cite{Gardenes2007}. However, under positive frequency-degree correlation 
this frustration is induced by an emergent frequency gap existing between hubs and their neighbors. 
Therefore, frustrating the path to synchronization in SF networks is tantamount to isolate the influence 
of the hubs in the system. In this way, the more connected the network is through the hubs, 
the more explosive the transition becomes once the hubs are isolated. 
We can quantify this effect by evaluating the node betweenness centrality, 
which computes the fraction of all shortest paths passing through each node of the network. 
Figure~\ref{fig5} shows the mean betweenness for the core made of the set of the first three higher-degree nodes 
(for the remaining nodes the betweenness does not change significantly). 
For the SF growing method, the network is more connected through the hubs than for CM static networks within 
the region of degree-degree correlation where explosive behavior is observed (see inset of Fig.~\ref{fig5}). 
Therefore, in the case of the CM there are more paths connecting the network that do not necessarily pass through the hubs, 
allowing progressive local synchronization and thus reducing the explosive character of the transition and 
the associated hysteresis width. 
This is of course due to the specific characteristics of the hubs in each network model: 
while a growing SF network starts from an all-to-all connected seed, for the static CM network the hubs are homogeneously 
distributed in the network, as their natural degree-degree correlations reveal: $r\simeq 0$ for growing SF networks and $r=-0.19$ for the CM ones.

Figure~\ref{fig4}, bottom panel, shows that the maximum of the hysteresis width is reached for a moderate 
increase of $r$ over the natural degree-degree correlation value of the original network. 
According to Ref.~\cite{Zou2014}, $\sigma_c$ increases with the degree of the main hub for uncorrelated SF networks in the limit of small mean degree networks, where the role of the hubs is certainly dominant. Therefore, we suggest that a small increase 
in the degree-degree correlation promotes the connectivity of the hubs leading to the emergence of a core 
with a larger effective degree, which increases the hysteresis width accordingly. However, further increase 
of the assortativity/dissassortativity enhances the modularity of the network, 
thus breaking the dominant role of hubs by over/under connecting them. This is reflected by the decrease of the core's 
betweenness for large positive and negative values of $r$ for both growing and static networks (see Fig.~\ref{fig5}). 
It should be noted that the overlapping of the betweenness centrality for growing and static networks only occurs 
for $r=-0.22$ and $r=0.16$, where there is no hysteresis width present in none of each models (bottom panel of Fig.~\ref{fig4}). This supports the theory that the structure of the core is mainly responsible for the hysteresis enhancement, 
since the betweenness is the same for growing and static networks only in the absence of hysteresis, that is, 
when large values of $|r|$ make both networks similar.

In summary, we have reported large scale simulations of the dynamics of
networked ensembles of phase oscillators whose interactions are mediated by a scale free topology of connections, and for which a
positive correlation exists between each oscillator's natural frequency and the corresponding node degree. 
Our results allow to conclude that
the further presence of a degree-degree mixing in the network structure has crucial
consequences on the nature of the phase transition accompanying the emergence of the
phase-coherent state of the network. In particular, we have shown that high levels of both positive and negative mixings consistently 
produce a second-order
phase transition, whereas moderate values of assortative mixing magnify the irreversible nature of ES in 
growing SF networks.
When reported to the fact that non-trivial forms of degree
correlation ubiquitously characterize, indeed, the structure of real world SF networks 
(with social networks typically displaying moderate levels of assortativity while
technological and biological networks exhibiting a disassortative mixing), our results may be of relevance for understanding why
real-world biological and technological networks organize themselves on topological structures that tend to avoid explosive synchronization phenomena
(which are there usually associated to pathological states of the networks), whereas social networks's topologies are actually favoring
the explosive and irreversible emergence of synchronous states.

The authors acknowledge financial support from the Spanish Ministerio de Econom\'ia y Competitividad 
under project FIS2012-38949-C03-01. Z.W. acknowledges the National Natural Science Foundation of
China (Grant No. 11005047).

\bibliography{references}

\end{document}